\documentclass[11pt,twoside]{article}
\usepackage{asp2014}

\aspSuppressVolSlug
\resetcounters

\begin{document}

\title{Asymmetric distribution of data products from WALLABY, an SKA precursor neutral hydrogen survey}

\author{Parra-Royon~M.$^1$, Austin~S.$^2$, Reynolds~T.$^3$ , Venkataraman~P.$^4$, Mendoza~MA.$^1$, Sanchéz-Expósito~S.$^1$, Garrido~J.$^1$, Kitaeff~S.$^2$ and Verdes-Montenegro~L.$^1$}

\affil{$^1$Instituto de Astrofísica de Andalucía (IAA-CSIC), Granada, Spain \email{mparra@iaa.es}}
\affil{$^2$CSIRO Space and Astronomy, PO Box 1130, Bentley WA 6102, Australia}
\affil{$^3$International Centre for Radio Astronomy Research (ICRAR), The University of Western Australia, 35 Stirling Hwy, Crawley, WA, 6009, Australia and ARC Centre of Excellence for All Sky Astrophysics in 3 Dimensions (ASTRO 3D)}
\affil{$^4$Canadian Initiative for Radio Astronomy Data Analysis, CIRADA, Canada.}


\paperauthor{Parra-Royon~M}{mparra@iaa.es}{0000-0002-6275-8242 }{Instituto de Astrofísica de Andalucía (IAA-CSIC)}{}{Granada}{}{1808}{Spain}
\paperauthor{Austin~S}{Austin.Shen@csiro.au}{}{CSIRO Space and Astronomy  }{}{Bentley}{}{WA 6102}{Australia}
\paperauthor{Reynolds~T}{tristan.reynolds@uwa.edu.au}{ 0000-0002-6606-5953}{International Centre for Radio Astronomy Research (ICRAR), The University of Western Australia, 35 Stirling Hwy, Crawley, WA, 6009, Australia and ARC Centre of Excellence for All Sky Astrophysics in 3 Dimensions (ASTRO 3D)}{}{Crawley}{}{WA 6009}{Australia}
\paperauthor{Venkataraman~P}{p.venkataraman@utoronto.ca}{0000-0003-0278-3398}{Canadian Initiative for Radio Astronomy Data Analysis, CIRADA.}{}{}{}{}{Canada}
\paperauthor{Mendoza~MA.}{amendoza@iaa.es}{0000-0003-3800-8668 }{Instituto de Astrofísica de Andalucía (IAA-CSIC)}{}{Granada}{}{1808}{Spain}
\paperauthor{Sanchéz-Expósito~S.}{sse@iaa.es}{0000-0002-7510-7633 }{Instituto de Astrofísica de Andalucía (IAA-CSIC)}{}{Granada}{}{1808}{Spain}
\paperauthor{Garrido~J.}{jgarrido@iaa.es}{ 0000-0002-6696-4772 }{Instituto de Astrofísica de Andalucía (IAA-CSIC)}{}{Granada}{}{1808}{Spain}
\paperauthor{Kitaeff~S.}{slava.kitaeff@monash.edu}{0000-0002-9690-9395 }{CSIRO Space and Astronomy  }{}{Bentley}{}{WA 6102}{Australia}
\paperauthor{Verdes-Montenegro~L.}{lourdes@iaa.es}{0000-0003-0156-6180 }{Instituto de Astrofísica de Andalucía (IAA-CSIC)}{}{Granada}{}{1808}{Spain}




  
\begin{abstract}

The Widefield ASKAP L-band Legacy All-sky Blind surveY (WALLABY) is a neutral hydrogen survey (HI) that is running on the Australian SKA Pathfinder (ASKAP), a precursor telescope for the Square Kilometre Array (SKA). The goal of WALLABY is to use ASKAP's powerful wide-field phased array feed technology to observe three quarters of the entire sky at the 21 cm neutral hydrogen line with an angular resolution of 30 arcseconds. 
Post-processing activities at the Australian SKA Regional Centre (AusSRC), Canadian Initiative for Radio Astronomy Data Analysis (CIRADA) and Spanish SKA Regional Centre prototype (SPSRC) will then produce publicly available advanced data products in the form of source catalogues, kinematic models and image cutouts, respectively. These advanced data products will be generated locally at each site and distributed across the network. 
Over the course of the full survey we expect to replicate data up to 10 MB per source detection, which could imply an ingestion of tens of GB to be consolidated in the other locations near real time. 
Here, we explore the use of an asymmetric database replication model and strategy, using PostgreSQL as the engine and Bucardo as the asynchronous replication service to enable robust multi-source pools operations with data products from WALLABY. 
This work would serve to evaluate this type of data distribution solution across globally distributed sites. Furthermore, a set of benchmarks have been developed to confirm that the deployed model is sufficient for future scalability and remote collaboration needs.
  
\end{abstract}

\section{Introduction}
\label{sec:intro}

The SKA Observatory (SKAO) will build the most sensitive telescope on the planet to address key questions in astrophysics, fundamental physics and astrobiology. The SKA will consist of two radio telescopes, one in Australia housing 131,072 dipole antennas, grouped into 512 stations, each with 256 antennas, and one array in South Africa with 197 dishes. In the process of developing this large science infrastructure, precursors and pathfinders, such as the South African MeerKAT and HERA, together with the Murchison Widefield Array (MWA) and the CSIRO's Australian SKA Pathfinder (ASKAP) are being put in place, providing SKA scientists with knowledge and technology to assist in the future design and deployment of the SKA. These facilities, and the SKA's future main telescope, will require the ability to process and manage Big Data, which will be a technical challenge for the coming years given its characteristics related to the 5 Vs (volume, velocity, variety, veracity and value). 

The data management, access and infrastructure model will be mainly distributed and this aspect plays a crucial role in the implementation of the SKA Regional Centre around the world. In this context, WALLABY \citet{koribalski2020WALLABY}, is expected to detect half a million galaxies in the local universe, producing about 75 PB of raw data each year. Derived from this, advanced data products resulting from multi-site post-processing workflows, such as those found in the AusSRC and SPSRC prototypes, or the CIRADA collaboration will be generated to take Level 6 data from the telescope and make them accessible to scientists. The data volume provide by WALLABY for 400-500 detections (in a period of 1-2 weeks) is approximately 5GB, which will be distributed among the different locations of the network in near real time. This scenario will require the replication and synchronisation of database tables across all sites. A multi-site data repository environment will be essential for efficient distribution management of the access, location and delivery of data. 

In addition, it is necessary to verify that the ingestion of these data products can be efficiently distributed to the rest, maintaining data consistency where each network site needs to improve independence from the source site. Different proposals for database replication, distribution and synchronisation have been developed to suit different scenarios. Here we explore a model that effectively address the problem with the distribution of WALLABY data products.

\begin{figure}[ht]
    \centering
    \includegraphics[width=\linewidth]{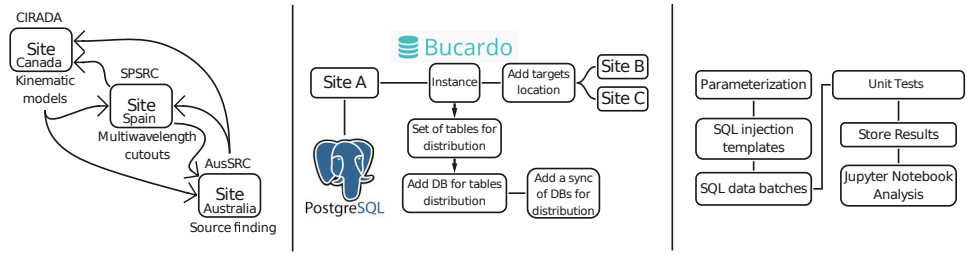}
    \caption{Structure of the data distribution of the production environment for data locations, sources and destinations (left image). Preparation of the data distribution environment for a Bucardo site (centre image). Design of the automated workflow for the generation, execution and results of the benchmark (right image).}
    \label{fig:fig1}
\end{figure}

\section{Problem and constraints}

To tackle this problem we need to distribute post-processing data products that are stored in databases and tables at local source sites. The sites are geographically separated and can generate data or be the destination of replicated data. Post-processing products are stored on local PostgreSQL \citep{thomas2017postgresql} Database Manager System from the sources location. These products can be defined in lightweight products, such as detections, pipeline executions, or kinematics models, and heavy data products such as cubes or cutouts. Lightweight data is about 10-100 KB in size per row, while product data is in the range of 2 to 10 MB per row, assuming a total of 400-500 detections and 5 GB per week. In our study, we will focus on the heavy data products, since they may be the bottleneck in the replication of this type of system. 



\section{Approaches}

In this context, to solve data synchronization, there are different tools and services to replicate data between locations using PostgreSQL as DBMS. 
Bearing in mind our knowledge base, a series of tests and validations\footnotemark{} taking into account the problem model and constraints described in the previous section, it is decided to deploy Bucardo, based on: a) it allows multi-master replication, b) it offers fine granularity in terms of data source selection, c) it offers multi-target and multi-source capabilities, d) it can be managed in a decentralised way and finally e) it is asynchronous.

\section{Data ingestion scenario, replication service and benchmark deployment}

Post-processing activities at the Australian SKA Regional Centre (AusSRC), Canadian Initiative for Radio Astronomy Data Analysis (CIRADA) and Spanish SKA Regional Centre prototype (SPSRC) will then produce publicly available advanced data products in the form of source catalogues, kinematic models and image cutouts, respectively. The idea is that each location will provide data sources that are replicated and synchronized with the rest, making the data globally available throughout the network.

In order to prepare this environment, an instance of the Bucardo service has been deployed at each of the location sites, simulating different sources and destinations of data products.  To inject data, a simulation with synthetic data has been performed from the Spanish SRC infrastructure \citep{garrido2021} taking into account the volume of data expected with WALLABY (see Section \ref{sec:intro}). With this information, a set of test scenarios have been designed, modeling data ingestion into the database through a benchmark detailed in our repository\footnotemark[\value{footnote}]. 

\footnotetext{\url{https://github.com/AusSRC/db_replication}}

\section{Results}

\begin{figure}[ht]
    \centering
    \includegraphics[width=\linewidth]{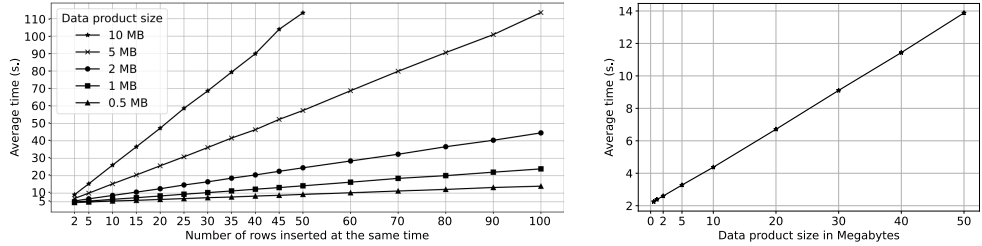}
    \caption{Evolution of the average consolidation time of different size data products in the SRCs network, for multiple row insertion sequences (left image). Note that for the 10MB data product we have found a limitation in PostreSQL when inserting products in large batches. Evolution of the average consolidation time of one data product over different data product sizes (right image).}
    \label{fig:fig2}
\end{figure}

After the execution of the tests a data set of results has been obtained. This dataset contains the following transactions: for each data product size [0.5, 1, 2, 5, 10] MBytes, insertion in sequences of [2, 5, 10, 15, 20, 25, 30, 35, 40, 45, 50, 60, 70, 80, 90, 100] rows at the same time, 5 times each, 
assuming a total of 400 transactions. Figure \ref{fig:fig2} shows a summary of the results where it is observed that the system is able to scale properly. A Jupyter Notebook is available in our repository to reproduce the results.

\section{Conclusions}

Replication of data products between locations can be a complicated task given the heterogeneity of the data and the large delays to overcome, which translate into data synchronisation consistency and performance issues respectively. In this work we design a data replication model and strategy that provides more than sufficient service levels for the distribution of WALLABY post-processing data products, following an asymmetric model.
This work will serve to evaluate this type of data distribution solution across globally distributed sites, confirming that the deployed model is sufficient for future scalability and remote collaboration needs.

\section*{Acknowledgements}


We acknowledge financial support from a) State Agency for Research of the Spanish Ministry of Science, Innovation and Universities through the "Center of Excellence Severo Ochoa" awarded to the Instituto de Astrofísica de Andalucía (SEV-2017-0709), b) grant RTI2018-096228-B-C31, c) grant EQC2019-005707-P , d) grant SOMM17-5208-IAA-2017, e)  budgetary line 28.06.000x.430.09 for the coordination of the participation in SKA-SPAIN, and e) grant 54A Scientific Research program D1113102E3.


\bibliography{X7-009}


\end{document}